\documentclass[]{llncs}
\usepackage{graphicx}
\usepackage{floatflt, wrapfig}
\usepackage{xcolor}
\usepackage{amsmath}
\usepackage{amssymb, amsbsy}
\usepackage{graphicx}
\graphicspath{ {images/} }
\usepackage{epstopdf}
\usepackage{color, import}
\usepackage{sectsty}
\usepackage{enumitem}
\usepackage{lineno}
\usepackage[hidelinks]{hyperref}
\usepackage{cite}
\usepackage[edges]{forest}
\usepackage[linesnumbered,vlined,ruled]{algorithm2e}
\SetAlFnt{\footnotesize}
\SetAlCapFnt{\small}
\SetAlCapNameFnt{\small}
\usepackage{setspace}
\usepackage[list=true]{subcaption}
\usepackage[title]{appendix}

\usepackage{array}

\usepackage{courier}

\usepackage{multirow, tabulary, tabularx}
\usepackage{array}
\usepackage{colortbl}
\usepackage{multicol}
\usepackage{hhline}

\usepackage{geometry}
 \newcolumntype{C}[1]{>{\centering\let\newline\\\arraybackslash\hspace{0pt}}m{#1}}

\begin{document}
\title{Positional Encoding by Robots with Non-Rigid Movements \thanks{This is the full version of the paper that was accepted in the 26th International Colloquium on Structural Information and Communication Complexity (SIROCCO 2019), July 1-4, 2019, L’Aquila, Italy.}}
%
%
\author{Kaustav Bose \and
Ranendu Adhikary \and
Manash Kumar Kundu \and
Buddhadeb Sau}
%
%
\institute{Department of Mathematics, Jadavpur University, Kolkata, India\\
\email{\{kaustavbose.rs, ranenduadhikary.rs, manashkrkundu.rs, buddhadeb.sau\}@jadavpuruniversity.in}
}
\maketitle                 

\begin{abstract}

Consider a set of autonomous computational entities, called \emph{robots}, operating inside a polygonal enclosure (possibly with holes), that have to perform some collaborative tasks. The boundary of the polygon obstructs both visibility and mobility of a robot. Since the polygon is initially unknown to the robots, the natural approach is to first explore and construct a map of the polygon. For this, the robots need an unlimited amount of persistent memory to store the snapshots taken from different points inside the polygon. However, it has been shown by Di Luna et al. [DISC 2017] that map construction can be done even by oblivious robots by employing a positional encoding strategy where a robot carefully positions itself inside the polygon to encode information in the binary representation of its distance from the closest polygon vertex. Of course, to execute this strategy, it is crucial for the robots to make accurate movements. In this paper, we address the question whether this technique can be implemented even when the movements of the robots are unpredictable in the sense that the robot can be stopped by the adversary during its movement before reaching its destination. However, there exists a constant $\delta > 0$, unknown to the robot, such that the robot can always reach its destination if it has to move by no more than $\delta$ amount. This model is known in literature as \emph{non-rigid} movement. We give a partial answer to the question in the affirmative by presenting a map construction algorithm for robots with non-rigid movement, but having $O(1)$ bits of persistent memory and ability to make circular moves.

\end{abstract}

\section{Introduction}

Distributed coordination of autonomous mobile robots has been extensively studied in literature in the last two decades. Fundamental problems like \textsc{Gathering} \cite{CieliebakFPS12,Flocchini08,Pagli15,agathangelou2013distributed}, \textsc{Pattern Formation} \cite{SuzukiY99,Cicerone17,Yamauchi14, Bose19} etc., have been studied in the setting where the robots are deployed in the plane with infinite extent and without any obstacles. Recently in \cite{Luna17}, \textsc{Meeting}, which is a simpler version of the \textsc{Gathering} problem, has been investigated for robots inside a polygonal enclosure containing polygonal obstacles, where their boundaries limit both visibility and mobility of a robot. This setting models many real life scenarios like moping robots inside a room, robots employed in factories or an art gallery etc. To solve the various distributed problems in this model, the robots may have to first explore and construct a map of the environment. For this, the robots need an unlimited amount of persistent memory. However, in \cite{Luna17}, it has been shown that map construction can be done even by oblivious robots with rigid movements, i.e., where a robot can accurately move by any distance. Their strategy is based on a positional encoding technique, where the robot carefully moves within the polygonal enclosure in such a way that their memory is implicitly encoded in its distance from the closest polygon vertex. In this paper, we show that this technique can be adapted to the non-rigid setting (where the movements of the robots can be interrupted by the adversary) as well, provided that the robot has a constant number of persistent bits and the ability to make circular moves. 

\subsection{Related Works}
The work that is closest to ours is \cite{Luna17}, where the \textsc{Meeting} problem was studied for a set of anonymous, oblivious and asynchronous robots in a polygon. The \textsc{Meeting} problem asks to design a distributed movement strategy for $n \geq 2$ robots,
so that eventually at least two of them come to see each other and become `mutually aware'. \textsc{Rendezvous} by two robots in polygons have been studied in \cite{CzyzowiczILP11,CzyzowiczPL12,CzyzowiczKP13,DieudonneP15}, where the two robots have to meet at a point or get arbitrarily close to each other. However, their model is significantly different from \cite{Luna17}. Another problem related to our map construction problem is that of constructing the visibility graph of a polygon by mobile robots \cite{Chalopin13a,Chalopin13b,Disser14,Chalopin15}.

\paragraph{Organization.} The paper is organized as follows. In Section \ref{model}, the basic model and relevant definitions are presented. In Section \ref{sec: overview}, a brief overview of the positional encoding technique is presented. Then in Section \ref{sec: main}, we describe our main algorithm. In Section \ref{conclu}, we briefly discuss how our algorithm and the techniques used in it can be used to solve some distributed coordination problems in polygonal environment.

\section{Model and Definitions}\label{model}

\paragraph{Polygon.} A \emph{polygon} $P$ is a non-empty, connected, and compact region in $\mathbb{R}^2$ whose boundary $\partial(P)$ is a set of finitely many disjoint simple closed polygonal chains. There is one connected component of $\partial(P)$, called the \emph{external boundary}, which encloses all others (if any), which are called \emph{holes}. Vertices and edges of a polygon can be defined in the standard way. $V(P)$ and $E(P)$ will respectively denote the set of vertices and edges of the polygon. For any two points $x, y \in P$, we say that \emph{$x$ and $y$ are visible to each other} if the line segment joining them lies in $P$, i.e., $\overline{xy} \subset P$. We shall assume that there is some global coordinate system, with respect to which, the coordinates of the polygon vertices are algebraic numbers.

\paragraph{Robot.} By a \emph{robot}, we mean an anonymous mobile computational entity modelled as a dimensionless point inside $P$. The robot is equipped with visibility sensors which allow it to observe its surroundings. Formally, a robot positioned at $x \in P$ can observe a point $y \in P$ if and only if $x$ and $y$ are visible to each other. The robot has movement capabilities that allow it to move inside the polygon along a straight line or a circular arc. The robot is endowed with $O(1)$ bits of persistent memory. This model is known in literature as $\textsc{FState}$ \cite{FlocchiniSVY16}, where the internal state of the robot can assume a finite number of `colors'. $\mathcal{S}$ will denote the set of all possible states of the robot.   

A robot, when active, operates according to the so-called \emph{LOOK-COMPUTE-MOVE} cycle. In each cycle, a previously idle robot wakes up and executes the following steps. In the LOOK phase, the robot takes a snapshot of the region of $P$ that it can currently see. The snapshot is
expressed in the local coordinate system of the robot having the origin at its current position. In the COMPUTE phase, based on the snapshot and its internal state, the robot performs computations according to a deterministic algorithm to decide 1) a destination point $y \in P$, 2) a trajectory to $y$ from its current location $x \in P$, which is either a straight line segment, or circular arc, 3) a state $s \in \mathcal{S}$. Then in the MOVE phase, the robot sets its internal state to $s$ and moves towards the point $y$ along the decided trajectory. When a robot transitions from one LCM cycle to the next, all of its local memory (past computations and snapshots) are completely erased, and only its internal state is retained. We shall assume that the local coordinate system of the robot is persistent, in the sense that its orientation, scale, and handedness are the same in each LCM cycle.

Depending on whether or not the adversary can stop a robot before it reaches its computed
destination, there are two movement models in literature, namely  \emph{rigid} and \emph{non-rigid}, respectively. In the rigid model, a robot is always able to reach its desired destination without any interruption. In the case of non-rigid
movements, there exists a constant $\delta > 0$, such that if the robot decides to move by an amount (path length) smaller
than $\delta$, then the robot will reach it; otherwise, it will move by at least $\delta$ amount. The value of $\delta$ is not known to the robot.

\paragraph{Geometric Definitions and Notations.}

Let $v$ be any vertex of $P$, and $u, w$ be its two adjacent vertices. We shall say that \emph{$u$ is the preceding vertex of $v$} and \emph{$w$ is the succeeding vertex of $v$} if one can reach from $\overline{vu}$ to $\overline{vw}$ by moving around $v$ (staying inside $P$) in the counterclockwise direction (according to the sense of handedness of the robot). For any vertex $p_i \in V(P)$, unless mentioned otherwise, $p_{i-1}$ and $p_{i+1}$ will respectively denote the vertices preceding and succeeding $p_i$.

\begin{floatingfigure}[r]{6.0cm}
    \fontsize{10pt}{10pt}\selectfont
    \def\svgwidth{0.3\textwidth}
\begingroup%
  \makeatletter%
  \providecommand\color[2][]{%
    \errmessage{(Inkscape) Color is used for the text in Inkscape, but the package 'color.sty' is not loaded}%
    \renewcommand\color[2][]{}%
  }%
  \providecommand\transparent[1]{%
    \errmessage{(Inkscape) Transparency is used (non-zero) for the text in Inkscape, but the package 'transparent.sty' is not loaded}%
    \renewcommand\transparent[1]{}%
  }%
  \providecommand\rotatebox[2]{#2}%
  \ifx\svgwidth\undefined%
    \setlength{\unitlength}{455.2bp}%
    \ifx\svgscale\undefined%
      \relax%
    \else%
      \setlength{\unitlength}{\unitlength * \real{\svgscale}}%
    \fi%
  \else%
    \setlength{\unitlength}{\svgwidth}%
  \fi%
  \global\let\svgwidth\undefined%
  \global\let\svgscale\undefined%
  \makeatother%
  \begin{picture}(1,0.5940246)%
    \put(0,0){\includegraphics[width=\unitlength]{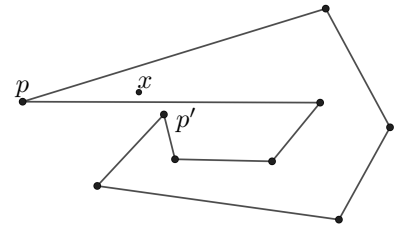}}%
    \put(0.01619519,0.34757365){\color[rgb]{0,0,0}\makebox(0,0)[lb]{\smash{$p$}}}%
    \put(0.42199329,0.25935871){\color[rgb]{0,0,0}\makebox(0,0)[lb]{\smash{$p'$}}}%
    \put(0.32428727,0.35968854){\color[rgb]{0,0,0}\makebox(0,0)[lb]{\smash{$x$}}}%
  \end{picture}%
\endgroup%

\caption{The polygon vertex closest to $x$ is $p'$, but it is not visibile to $x$. Its closest visible vertex is $p$.}
\label{fig: vis1}
\end{floatingfigure}

For a set $X = \{x_1,x_2,\ldots,x_n\}$ of distinct points in $\mathbb{R}^2$, $n \geq 2$, the \emph{Voronoi region} of any $x_i \in X$, denoted by $Vor_X(x_i)$ or simply $Vor(x_i)$, is the set of all points in $\mathbb{R}^2$ which are closer to $x_i$ than any other point in $X$, that is, $Vor_X(x_i) = \{y \in \mathbb{R}^2 \mid d(y, x_i) \leq d(y, x_j), \forall i \neq j\}.$ Points shared by two Voronoi regions $Vor_X(x_i)$ and $Vor_X(x_j)$ constitute the \emph{Voronoi edge} defined by $x_i$ and $x_j$. Similarly, we can define Voronoi regions for a set $L = \{l_1,l_2,\ldots,l_n\}$,$n \geq 2$ of straight line segments (any two of which can intersect only at their endpoints). We will define the Voronoi region of $l_i \in L$ as $LVor_L(l_i) = \{y \in \mathbb{R}^2 \mid d(y, l_i) \leq d(y, l_j), \forall i \neq j\}$ where $d(y, l_k) =$ \emph{Inf}$\{d(y, z) \mid z \in l_k\}$. In the context for our problem, there is a minor technical issue that needs to be addressed. For a polygon $P$, the polygon edge closest to a point $x \in P$ is of course visible to it. But the vertex closest to $x$ may not be visible from $x$ (See Fig. \ref{fig: vis1}). In the remainder of the paper, unless mentioned otherwise, whenever we say `closest vertex', it should be understood as `closest visible vertex'. We will also define the \emph{polygon Voronoi region} of a vertex $p_i$, denoted by $PVor_P(p_i)$, as the set of points  $x \in P$ such that $p_i$ is visible to $x$ and $p_i$ is closer to $x$ than any other vertex visible from $x$. $Vor_{V(P)}(p_i)$ or $Vor_P(p_i)$ will denote the usual Voronoi region of $p_i$ for the set $V(P)$.

For any point $x$, and any real number $r > 0$, $D(x,r)$ denotes the closed disc $\{y \in \mathbb{R}^2 \mid d(y,x) \leq r\}$. For any three points $c, y, z$ such that $d(y,c) = d(z,c)$, we shall denote by $arc(y,z,c)$, the circular arc centered at $c$ drawn from $y$ to $z$ in counterclockwise direction. Also, $arc(y,\theta,c)$ will denote the circular arc $arc(y,z,c)$ where $\angle ycz = \theta$. A point $x \in P$ is said to be \emph{properly close} to $p_i \in V(P)$, if for any point $z \in arc(x,y,p_i)$, where $y \in \overline{p_ip_{i+1}}$ with $d(y,p_i) = d(x,p_i)$, the following holds: 1) $z \in PVor_{V(P)}(p_i)$ and 2) $p_{i+1}$ is visible from $z$.

We can define a coordinate system by any ordered pair of distinct points in the polygon. The coordinate system defined by $(u,v)$ will be the coordinate system with origin at $u$, $\overrightarrow{uv}$ as the positive $X$-axis, $d(u,v)$ as the unit distance and the positive $Y$-axis according to the chirality or handedness of the robot.

\section{A Brief Overview of the Positional Encoding Technique}\label{sec: overview}


\paragraph{Computational Model.}

We assume that each robot internally runs a \emph{Blum-Shub-Smale machine} \cite{Blum97} extended with a square-root primitive. A Blum-Shub-Smale machine  is a random-access machine whose registers can store arbitrary real numbers and can operate directly on them. Its computational primitives are the four basic arithmetic operations on real numbers, and it can test whether a real number is positive. Each of these operations takes one unit of time. Depending on the application, it is also customary to extend the basic model with additional 
primitives, such as root extractions, trigonometric functions, etc. In our case, we only require the square-root primitive that will be needed in geometric computations.

\paragraph{Encoding Algebraic Reals.}

 Consider an algebraic real number $\alpha$. The minimal polynomial of $\alpha$ over $\mathbb{Q}$ is the unique monic polynomial in $\mathbb{Q}[x]$ of least degree which has $\alpha$ as a root. Let $\mathfrak{m}(x) = x^n + a_{n-1}x^{n-1} + \ldots + a_1x + a_0 \in \mathbb{Q}[x]$ be the minimal polynomial of $\alpha$ over $\mathbb{Q}$. Now $\mathfrak{m}$ has $n$ complex roots. However, the real roots can be arranged in ascending order. So, let $\alpha$ be the $i$th real root of  $\mathfrak{m}$. Then $\alpha$ can be uniquely represented by $(n,i,a_{n-1},\dots,a_0)$. Now any rational number $(-1)^s\frac{p}{q}$, with $p, q > 0$, $s \in \{0,1\}$, can represented as a 3-tuple of non-negative integers as $(s,p,q) \in \mathbb{Z}^3_{\geq 0}$. Thus $\alpha$ can be represented by an array of $3n+2$ non-negative integers. We can represent each non-negative integer $m$ as the bit string $0^m1$. Let us denote by $\beta(\alpha)$, the bit string obtained by concatenating the bit strings of the $3n+2$ non-negative integers.
Now for any non-negative integer $\lambda$, let $r(\alpha, \lambda) < 1$ be the real number whose (usual) binary representation is $0.0^{\lambda}1\beta(\alpha)$.  We  shall say that \emph{$r(\alpha, \lambda)$ encodes $\alpha$}.


\begin{lemma}\label{scale}
 If $0 < d < 1$ be a real number such that $d = r(\alpha, \lambda)$, for some algebraic real $\alpha$ and non-negative integer $\lambda$, then $\frac{d}{2} = r(\alpha, \lambda + 1)$. Therefore, $\frac{d}{2^k} = r(\alpha, \lambda + k)$, for any integer $k \geq 1$.
\end{lemma}

\paragraph{Computing the Code.}

Suppose a basic Blum-Shub-Smale machine has an algebraic number $\alpha$ stored in its register and it has to construct its code $\beta(\alpha)$. The machine will generate all finite sequences of bits in lexicographic order. For each sequence, it will check if it is a
well-formed code of an algebraic number; if it is, it will extract the coefficients of the polynomial $\mathfrak{q}$
from it. Then it computes $\mathfrak{q}(\alpha)$. Since $\alpha$ is algebraic, eventually a polynomial $\mathfrak{q}$ is found such that $\mathfrak{q}(\alpha) = 0$. Since $\mathfrak{q}$ must be a multiple of the minimal polynomial $\mathfrak{m}$ of $\alpha$, we can determine it by  
finding its irreducible factor that has $\alpha$ as a root. Then Sturm's theorem \cite{cohen2013} can be applied to find out how many real roots of the minimal polynomial are smaller than $\alpha$. Thus we have obtained all that are required to encode $\alpha$. 


\paragraph{Computations on the Implicit Form.}

 Once a number is encoded in this form, we cannot necessarily retrieve it in finite time. But we can  approximate it arbitrarily well, for instance via Sturm's theorem. However, we can do Turing-computable bit manipulations on this implicit form to compute all kinds of common functions (e.g. basic arithmetic operations, root extractions of any degree etc.) on the algebraic number without decoding its explicit form. 


\paragraph{Encoding Snapshots.}

A snapshot taken by a robot contains the visible portion of the polygon $P$, which is basically a union of line segments, each of which being a sub-segment of an edge of $P$. So, a snapshot can be represented as an array of real numbers, say $S = (x_1, y_1, x'_1, y'_1, x_2, y_2, x'_2, y'_2, \ldots)$, where $(x_i, y_i)$ and $(x'_i, y'_i)$ are the endpoints of the $i$th visible segment of $\partial(P)$. Note that none of these points is necessarily a vertex of $P$. 

We have discussed how to compute the code of a single algebraic number. Now we describe how we can encode a snapshot of $P$ with algebraic vertices taken from a point $x \in P$. The vertices of $P$ have algebraic coordinate with respect to some global coordinate system. Of course, the vertices may not have algebraic coordinates in the local coordinate system of the robot. Let $\Phi_x$ be the transformation from the global coordinate system to the local coordinate system of the robot. Note that $x$ is not necessarily an algebraic point, and the parameters of $\Phi_x$ are not necessarily algebraic numbers either. Therefore, the coordinates and the distances between vertices of $\Phi_x(P)$ may not be algebraic. However, all the ratios of the distances are algebraic, as $\Phi_x$, being a similarity transformation, preserves ratios between segment lengths. Then it follows that if the robot picks two visible vertices of $\Phi_x(P)$, say $v$ and $v'$, and transforms all the visible vertices of $\Phi_x(P)$ in the coordinate system $(v, v')$, then they will have algebraic coordinates. Then they can be encoded by a basic Blum-Shub-Smale machine as we discussed earlier. However, recall that a snapshot taken from $x$ may not contain only vertices of $\Phi_x(P)$. We can identify the potentially non-vertex endpoints by a basic Blum-Shub-Smale machine, as a non-vertex point $(x_j, y_j) \in S$ is necessarily of the form $(x_j, y_j) = c(x_i,y_i), c > 1$ for some visible polygon vertex $(x_i,y_i)$. These potentially non-vertex endpoints will be simply marked with an `undefined' flag in the snapshot. The robot will pick two `defined' points in the snapshot for the coordinate transformation. The coordinates of the `defined' points of $S$ will be transformed as discussed earlier, and each `undefined' point will be simply replaced with a $(0, 0)$ or any algebraic point of our choice along with the `undefined' flag. Then these coordinates can be encoded into a finite bit string, and then they can be concatenated into a single code for the entire snapshot. We can similarly encode multiple snapshots into a single bit string. Along with the snapshots, we can also pack as many other finitely described elements as we want.

\paragraph{Positional Encoding.}

Suppose that $\beta$ is the code or bit string of the information that the robot wants to encode. Let $d$ be a real number that encodes it, i.e., the binary representation of $d$ is $0.0^{\lambda}1\beta(\alpha)$ for some non-negative integer $\lambda$. The robot will encode the information by positioning itself in the polygon in such a way that its distance from the closest polygon vertex is $d$ (according to its local coordinate system). From Lemma \ref{scale}, it follows that the robot can encode the same information by placing itself at a distance $\frac{d}{2^k}$ from the vertex for any integer $k \geq 1$. This `scalability' property allows the robot to get arbitrarily close to the
vertex without losing information.

\section{The Algorithm} \label{sec: main}

 In \cite{Luna17}, the memory of a robot is encoded in the distance from its closest polygon vertex. Obviously, the robot needs rigid movements to accurately position itself at a point whose distance from the particular vertex correctly encodes the memory. In the non-rigid setting, we need some additional options where we can encode our memory. In particular, apart from the distance from some particular vertex, we shall also encode the memory in the tangent of the angle that the robot makes with an edge or a diagonal, at some vertex. Note that computing the tangent of an angle in the current snapshot, and also computing a destination point, so that the tangent of the angle it makes with a line, at a vertex, is some given value, involves only the basic arithmetic operations and square root extraction. In the remainder of the paper, whenever we say that the memory is encoded in some angle $\alpha$, it is to be understood that the memory is encoded by the real number $tan (\alpha)$. Notice that since $tan (\alpha)$ monotonically tends to 0, as $\alpha < \frac{\pi}{2}$ tends to 0, we can use the scalability property of the encoding scheme to encode the memory in an angle as small as we want.

The persistent bits or the internal states are used so that each time a robot wakes up, it knows `where' its memory is encoded and which coordinate system the snapshots in the memory are expressed in. In each case, the robot also sets a particular polygon vertex, that is visible to it, as its \emph{virtual vertex}. A summary of this is provided in Table \ref{tab}.

\begin{table}
\centering
\begin{tabular}{|C{1cm}|C{5cm}|C{5.5cm}|C{4cm}|}
    \hline
    \multicolumn{4}{|c|}{\textbf{For any robot $r$ at a point $x$ inside the polygon $P$}}\\
    \hline
    \multirow{2}{*}{\textbf{State}} &\multirow{2}{*}{\textbf{Virtual Vertex}} &\multicolumn{2}{c|}{\textbf{Memory}}\\
   \hhline{|~|~|-|-|} &{}  &\textbf{encoded in} 	&\multicolumn{1}{c|}{\textbf{expressed in the coordinate system}}\\
    \hhline{|-|-|-|-|}
    $s_1$	&$p_i =$ the closest visible vertex	&$d(x,p_i)$	&\multicolumn{1}{c|}{$(p_i,p_{i+1})$}\\
    \hline
    $s_2$	&$p_i =$ the closest visible vertex	&$d(x,p_i)$	&\multicolumn{1}{c|}{$(p_{i-1},p_i)$}\\
    \hline
    $s_3$	& $p_a =$ the nearer endpoint of the closest boundary segment, say $\overline{p_ip_{i+1}}$, $a \in \{i,i+1\}$
    &$tan (\angle xp_ap_b)$, where $p_b$ is the other endpoint of $\overline{p_ip_{i+1}}$	&\multicolumn{1}{c|}{$(p_i,p_{i+1})$}\\
    \hline
    $s_4$	&$p_i =$ the closest visible vertex	&$tan (\angle xp_ip_{i-1} - \frac{\pi}{2})$	&\multicolumn{1}{c|}{$(p_{i-1},p_i)$}\\
    \hline
    $s_5$	&$p_i =$ the closest visible vertex	&$tan (\pi - \angle xp_ip_{i+1})$	&\multicolumn{1}{c|}{$(p_i,p_{i+1})$}\\
    \hline
    $s_6$	&$p_i =$ the closest visible vertex	&$tan (\angle xp_iO)$, where ${p_iO}$ is the angle bisector of $\angle p_{i-1}p_ip_{i+1}$	&\multicolumn{1}{c|}{$(p_{i-1},p_i)$}\\
    \hline
    $s_7$	&$p_i =$ the closest visible vertex	&$tan (\angle xp_i,p_j)$, where either $x$ lies on the interior of the Voronoi edge ${PVor}(p_i)\cap PVor(p_j)$ or $\overrightarrow{p_ix}$ intersects ${PVor}(p_i)\cap PVor(p_j)$ first	&\multicolumn{1}{c|}{$(p_i,p_{j})$ or $(p_j,p_{i})$}\\
    \hline
\end{tabular}\caption{The virtual vertex and encoded memory of the robot, corresponding to its internal state.}\label{tab}
\end{table}

Our map construction algorithm is similar to the one presented in \cite{Luna17}. The robot will keep exploring new vertices (but not touching it), and near each vertex, it will take a new snapshot and encode it, merging with the old snapshots.  As it explores, it keeps track of the vertices that it has seen but not yet visited. Whenever it reaches a new connected component of the boundary, it explores it entirely in the counterclockwise direction (i.e., by moving from a vertex to its succeeding vertex). After exploring a connected component for the first time, it will take a second tour of it, in the same direction. After completely exploring a previously unexplored connected component, it will choose an unvisited vertex of a different component and move to it via a suitable path. The robot repeats this until there are no unvisited vertices recorded in its encoded memory. 

Implementation of this strategy in the non-rigid setting is based on four basic techniques that we shall discuss in Section \ref{teq}. From there follows the main result of the paper presented in Section \ref{result}.

\subsection{Four Basic Techniques}\label{teq}

%
%
%
%
%

\subsubsection{Moving from One Virtual Vertex to Another in the Same Connected Component of the Boundary}\label{teq1}

\begin{figure}[h!]
\subcaptionbox[Short Subcaption]{
       \label{fig: linearA}
}
[
    0.49\textwidth 
]
{
    \fontsize{10pt}{10pt}\selectfont
    \def\svgwidth{0.49\textwidth}
    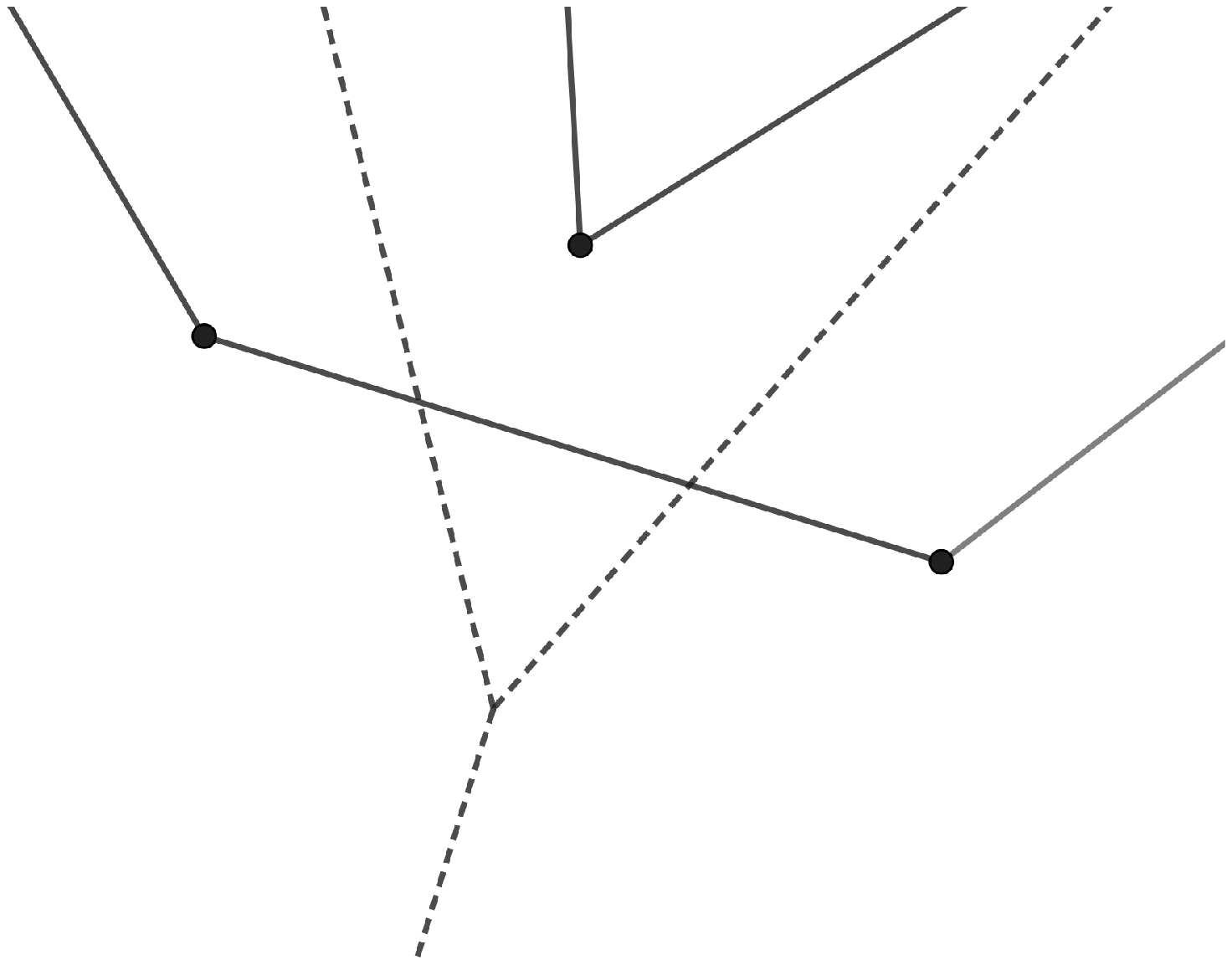
}
\hfill
\subcaptionbox[Short Subcaption]{
     \label{fig: linearB}
}
[
    0.49\textwidth 
]
{
    \fontsize{10pt}{10pt}\selectfont
    \def\svgwidth{0.49\textwidth}
    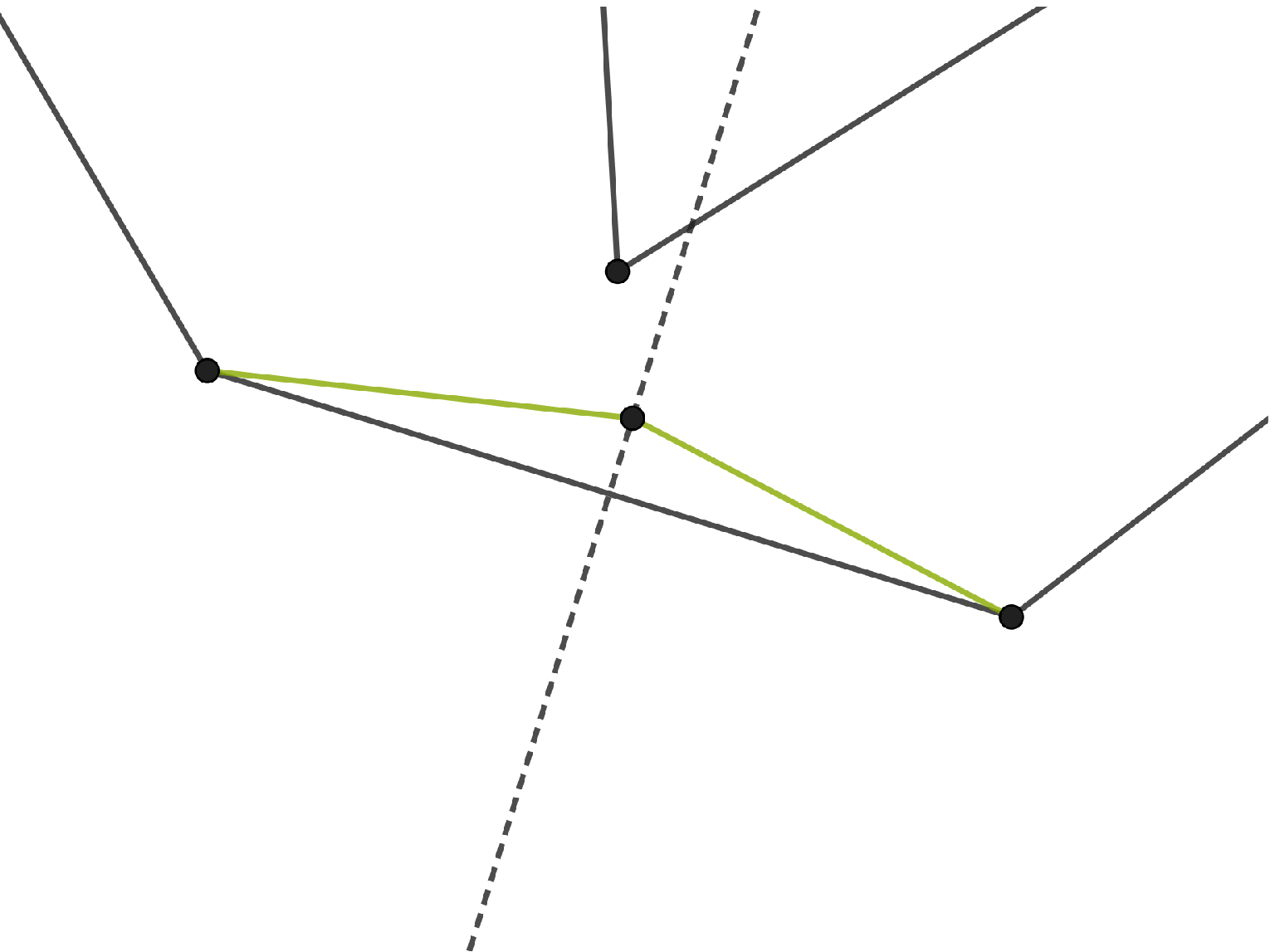
}
\caption[Short Caption]{a) If a robot moves from $p_i$ towards $p_{i+1}$, it has to pass through the Voronoi region of $p_j$ . b) The robot will move along the path $\mathcal{P}(p_i,p_{i+1},\alpha)$ drawn in green.}
\label{}
\end{figure}

 Suppose that $p_{i}$ is the virtual vertex of the robot $r$ with internal state $s_1$ (i.e., $p_i$ is the vertex closest to $r$), and it has to approach the succeeding vertex $p_{i+1}$. If $r$ had rigid movements, it could have simply moved to a point suitably close to $p_{i+1}$ in one go, without any interruption. But since $r$ has non-rigid movements, it can be stopped multiple times during its journey. Now consider the situation shown in Fig. \ref{fig: linearA}. To move towards $p_{i+1}$ via any path, the robot has to pass through the Voronoi region of $p_j$. Hence, if $r$ is stopped by the adversary while it is in the interior of $PVor_{V(P)}(p_j)$, it will set $p_j$ as its virtual vertex. To resolve this, the robot will change its state to $s_3$ before moving. When its state is $s_3$, to set the virtual vertex, it considers the closest boundary segment, instead of the closest vertex. The endpoint of its closest boundary segment that is closer to it, is set as the virtual vertex. In case of a tie, any one of the endpoints can be chosen as the virtual vertex. The robot will move along a path as shown in Fig. \ref{fig: linearB}. Such a path can be defined by a tuple $(p_i,p_{i+1},\alpha)$, where the path consists of two linear segments $\overline{p_{i}q}$ and $\overline{qp_{i+1}}$ of equal length with $\angle qp_ip_{i+1} = \angle qp_{i+1}p_i = \alpha$ and $q$ lying on the perpendicular bisector of $\overline{p_ip_{i+1}}$. We shall denote the path as $\mathcal{P}(p_i,p_{i+1},\alpha)$. The path should be chosen in such a way that any point on the path is closer to the boundary segment $\overline{p_{i}p_{i+1}}$ than any other point of $\partial(P)$. In other words, $\mathcal{P}(p_i,p_{i+1},\alpha)$ should be  inside $LVor_{E(P)}(\overline{p_ip_{i+1}})$.

Now let us describe our strategy more formally. Suppose that a robot $r$ is at a point $x$ inside the polygon $P$, such that the following are true.

\begin{description}

 \item [A1] $r.state = s_1$.
 \item [A2] $x$ is properly close to the vertex $p_i$.
\end{description}

Since $r.state = s_1$, $p_i$ is the virtual vertex of $r$, and its memory is encoded in the distance $d(x,p_i)$ and expressed in the coordinate system defined by $(p_{i},p_{i+1})$. Since $r$ is properly close to $p_i$, if $r$ moves around $p_i$ along a circular arc in counterclockwise direction (i.e., keeping its distance from $p_i$ fixed), $p_i$ will remain its virtual vertex and also, all of $\overline{p_{i}p_{i+1}}$ will remain visible to it. So, $r$ will move  around $p_i$ in counterclockwise direction to move to a point $x'$ such that the following conditions are satisfied.
\begin{description}
 \item [B1] the data encoded by $\alpha = \angle x'p_ip_{i+1}$ is same as the data encoded by $d(x',p_i) = d(x,p_i)$ both expressed in the coordinate system  $(p_{i},p_{i+1})$.
 
 \item [B2] the path $\mathcal{P}(p_i,p_{i+1},\alpha)$ is inside $LVor_{E(P)}(\overline{p_ip_{i+1}})$. 
\end{description}
%

%
%

After reaching such a point $x'$, $r$ will change its state to $s_3$. It will then follow the path $\mathcal{P}(p_i,p_{i+1},\alpha)$, where $\alpha = \angle x'p_ip_{i+1}$, to move towards $p_{i+1}$. However, note that $r$ had decided that $\mathcal{P}(p_i,p_{i+1},\alpha) \subset LVor_{E(P)}(\overline{p_ip_{i+1}})$, based on its view from $x'$. Therefore, it might happen that during its journey along the path $\mathcal{P}(p_i,p_{i+1},\alpha)$, it takes a snapshot at some point $y \in \mathcal{P}(p_i,p_{i+1},\alpha)$ and finds that $\overline{p_{i}p_{i+1}}$ is not its closest boundary segment. We can easily show that this is not possible.

\begin{figure}[h]
\centering
    \fontsize{10pt}{10pt}\selectfont
    \def\svgwidth{0.4\textwidth}
    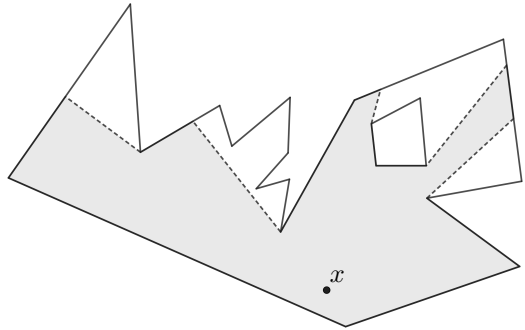

\caption{The perception of the polygon when seen from the point $x$.}
\label{perception}
\end{figure}

We have to show that if $y \in \mathcal{P}(p_i,p_{i+1},\alpha)$ and $z \in \partial(P) \setminus \overline{p_{i}p_{i+1}}$, such that $z$ is visible from $y$, then $d(\overline{p_{i}p_{i+1}}, y) < d(y,z)$. Let us denote by $P'$, the perception of $r$ of the polygon $P$ based on the snapshot taken at $x'$ (See Fig. \ref{perception}). $LVor(\overline{p_ip_{i+1}})$ according to $r$ at $x'$ is based on $P'$, and hence, can differ from the actual Voronoi region of $\overline{p_{i}p_{i+1}}$. Therefore, it is better to denote it by $LVor_{P'}(\overline{p_ip_{i+1}})$.
 It is easy to see that either $z \in \partial(P')$ or $z \in \mathbb{R}^2 \setminus P'$. If $z \in \partial(P')$, then clearly $d(\overline{p_{i}p_{i+1}}, y) < d(y,z)$ as $y \in LVor_{P'}(\overline{p_ip_{i+1}})$. Now consider the case where $z \in \mathbb{R}^2 \setminus P'$. Let $z'$ be the point where $\overline{yz}$ intersects $\partial(P')$. Again, $d(\overline{p_{i}p_{i+1}}, y) < d(y,z')$ as $y \in LVor_{P'}(\overline{p_ip_{i+1}})$. Therefore, $d(\overline{p_{i}p_{i+1}}, y) < d(y,z') < d(y,z') + d(z',z) = d(y,z)$ $\Rightarrow  d(\overline{p_{i}p_{i+1}}, y) < d(y,z)$.

We have not yet specified how close $r$ should get to $p_{i+1}$. Our objective is to get close to $p_{i+1}$, take a new snapshot and encode the new snapshot (merged with the older ones) in its distance from $p_{i+1}$. We want these snapshots to be expressed in the coordinate system defined by $(p_{i+1},p_{i+2})$, where $p_{i+2}$ is the vertex succeeding $p_{i+1}$. But in order to do that, $p_{i+2}$ should be visible to the robot. Notice that if some portion of $\overline{p_{i+1}p_{i+2}} \setminus \{p_{i+1}\}$ is visible to $r$, then it will be able to see all points of $\overline{p_{i+1}p_{i+2}}$ if it goes close enough to $p_{i+1}$. However, if $\overline{p_{i+1}p_{i+2}} \setminus \{p_{i+1}\}$ is completely invisible to $r$, the segment $\overline{p_{i+1}p_{i+2}}$ will never be completely visible to it, no matter how close it gets to $p_{i+1}$. In this section, we will only discuss the first case. The later case is more complex and will be discussed in the next section. 


So, consider the case where some portion of $\overline{p_{i+1}p_{i+2}} \setminus \{p_{i+1}\}$ is visible to $r$. In this case, $r$ will move to a point $x''$ that is close enough to $p_{i+1}$ so that the following conditions are satisfied.

\begin{description}
 \item [C1] $x''$ is properly close to $p_{i+1}$.
 \item [C2] $d(x'',p_{i+1})$ is  encoding the old snapshots merged with its current view (newly discovered vertices) all expressed in the coordinate system defined by $(p_{i+1},p_{i+2})$. 
\end{description}

The robot will first move close enough to $p_{i+1}$, say at $x'''$, so that the first condition is satisfied (See Fig.\ref{fig: properly}), i.e., $x'''$ is properly close to $p_{i+1}$. Then the robot decides to further move towards $p_{i+1}$ to a suitable point $x''$ in order to fulfill the last condition. There are two ways it may fail to achieve this. 

\begin{enumerate}
 \item If $d(x'',x''') > \delta$, the adversary can stop it at some point $x''''$ in between. However, the old snapshots are still available as it is encoded in $\angle x''''p_{i+1}p_i = \angle x'''p_{i+1}p_i$. So, $r$ can identify that it has failed to reach its destination. Then it will recompute the destination and move towards it.
 
 \item Even if it reaches $x''$, a new vertex may be discovered which is not present in the data encoded in $d(p_{i+1}, x'')$. Therefore, $r$ will again recompute a destination so that the newly discovered vertices are encoded (along with the old data).
\end{enumerate}

From the existence of $\delta > 0$ and the fact that the polygon has finitely many vertices, it follows that $r$ can eventually reach a point $x''$ where it finds that $d(x'',p_{i+1})$ encodes precisely the data encoded by $\angle x''p_{i+1}p_i$ merged with the new vertices of the polygon that are visible from $x''$. Observe that the visibility of both $\overline{p_{i+1}p_{i}}$ and $\overline{p_{i+1}p_{i+2}}$ are crucial at any point during this process. This is because the robot has to transform the data encoded in $\angle x''p_{i+1}p_i$ from the coordinate system $(p_{i},p_{i+1})$ to $(p_{i+1},p_{i+2})$. When all three $p_i,p_{i+1},p_{i+2}$ are visible, the robot knows their exact positions and hence, it can perform this conversion, which is computable by a rational function, on (the implicit form of) the old snapshots. When the conditions C1, C2 are achieved, $r$ will change its state to $s_1$. Clearly we are back to the situation where A1, A2 holds ($p_i$ to be replaced with $p_{i+1}$), and hence $r$ can now move to $p_{i+1}$ in the same manner.


 \begin{figure}[h!]
\subcaptionbox[Short Subcaption]{
       \label{fig: properly1}
}
[
    0.45\textwidth 
]
{
    \fontsize{10pt}{10pt}\selectfont
    \def\svgwidth{0.45\textwidth}
    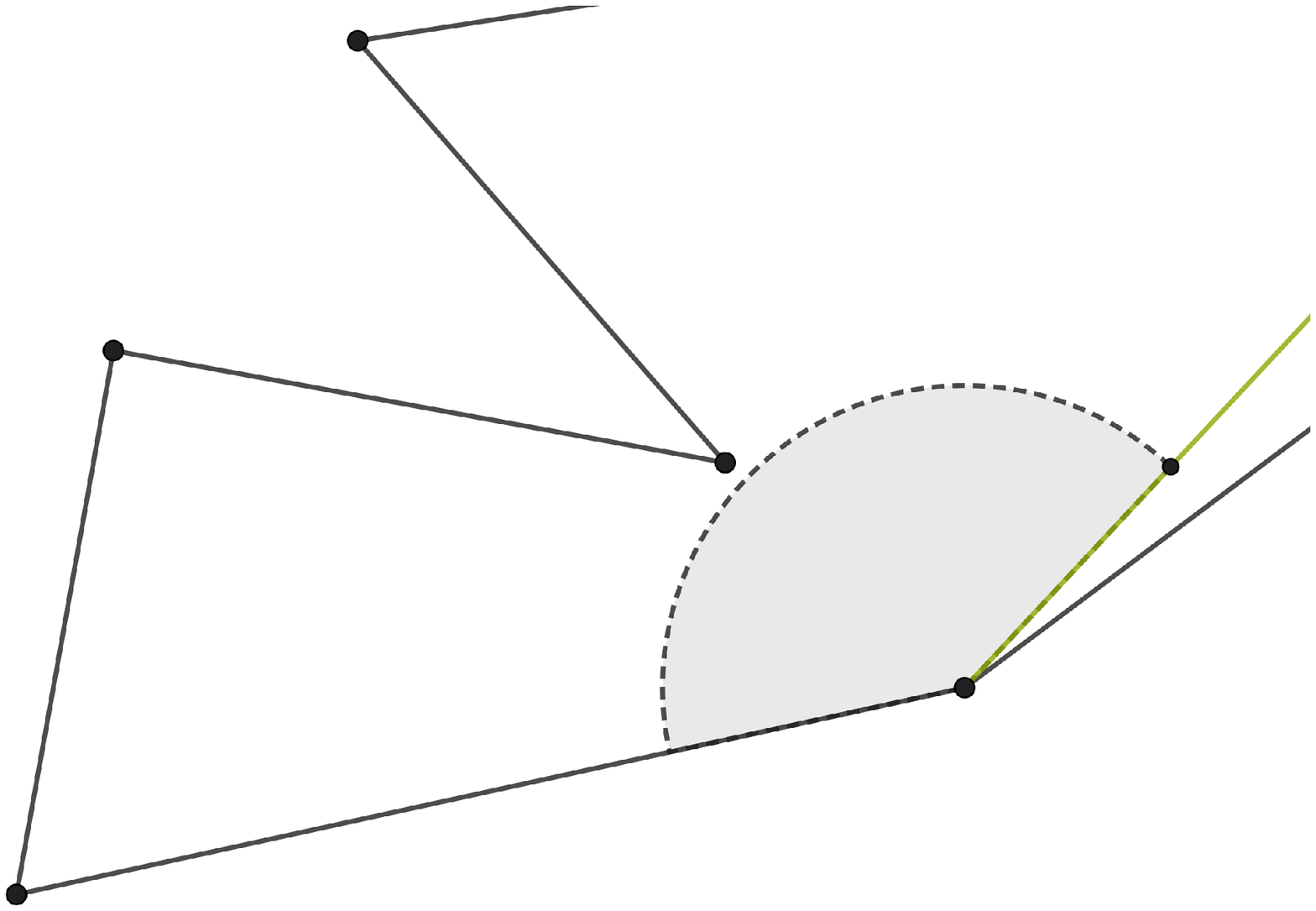
}
\hfill
\subcaptionbox[Short Subcaption]{
     \label{fig: properly}
}
[
    0.54\textwidth 
]
{
    \fontsize{10pt}{10pt}\selectfont
    \def\svgwidth{0.54\textwidth}
    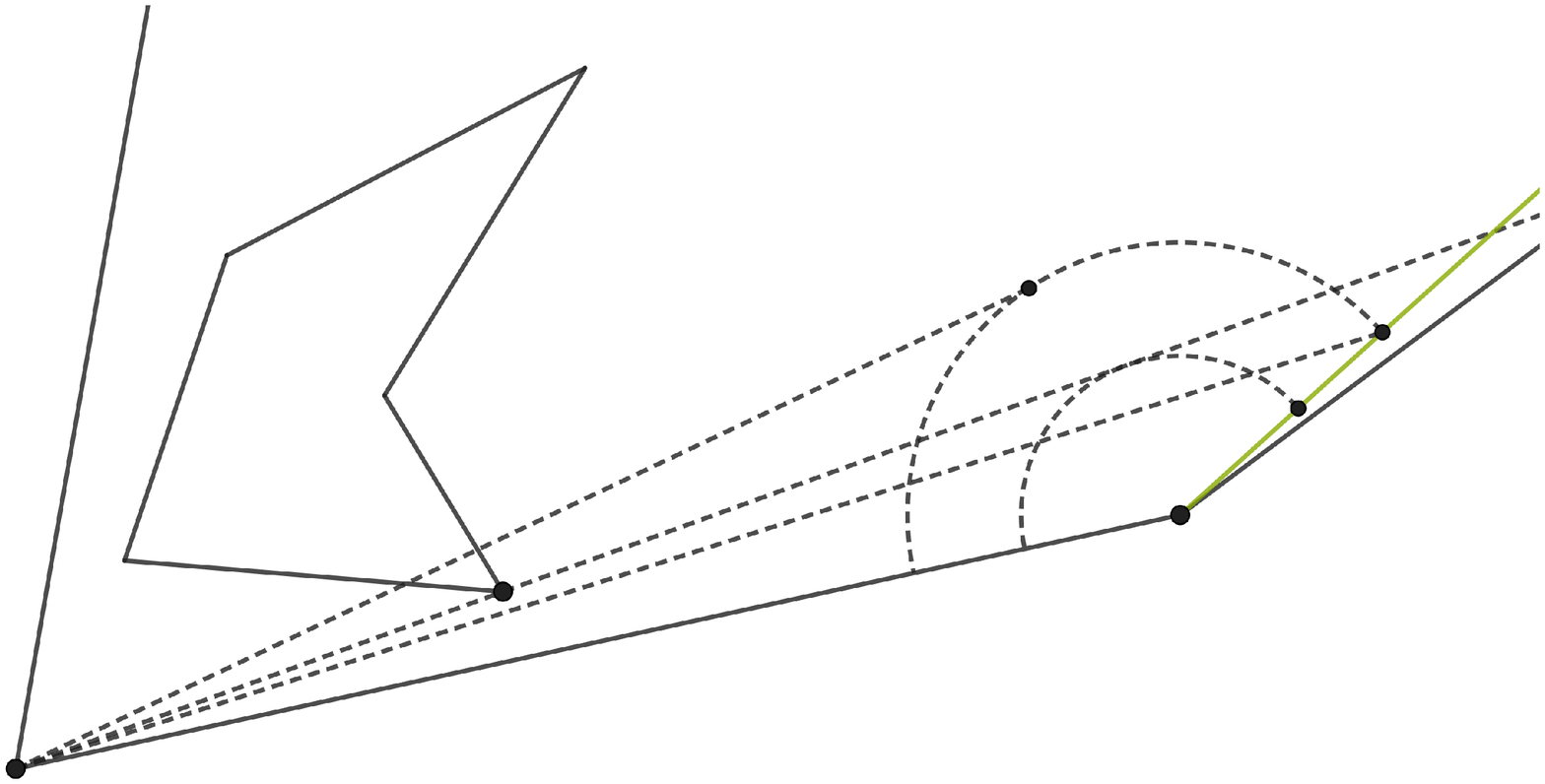
}
\caption[Short Caption]{a) The shaded circular sector of radius $d = d(p_{i+1},y)$ intersects no vertex other than $p_{i+1}$. Any point on $\overrightarrow{p_{i+1}y}$ less than $\frac{d}{2}$ distance away from $p_{i+1}$ satisfies the first condition of proper closeness to $p_{i+1}$. b) Any point on the interior of $\overline{p_{i+1}y}$ satisfies the second condition of proper closeness to $p_{i+1}$.}
\label{fig: properly}
\end{figure}

\subsubsection{Discovering the Succeeding Vertex and Encoding a New Snapshot}\label{teq2}

\begin{figure}[h!]
\subcaptionbox[Short Subcaption]{
       \label{fig: stage2_robot_order}
}
[
    0.49\textwidth 
]
{
    \fontsize{10pt}{10pt}\selectfont
    \def\svgwidth{0.49\textwidth}
    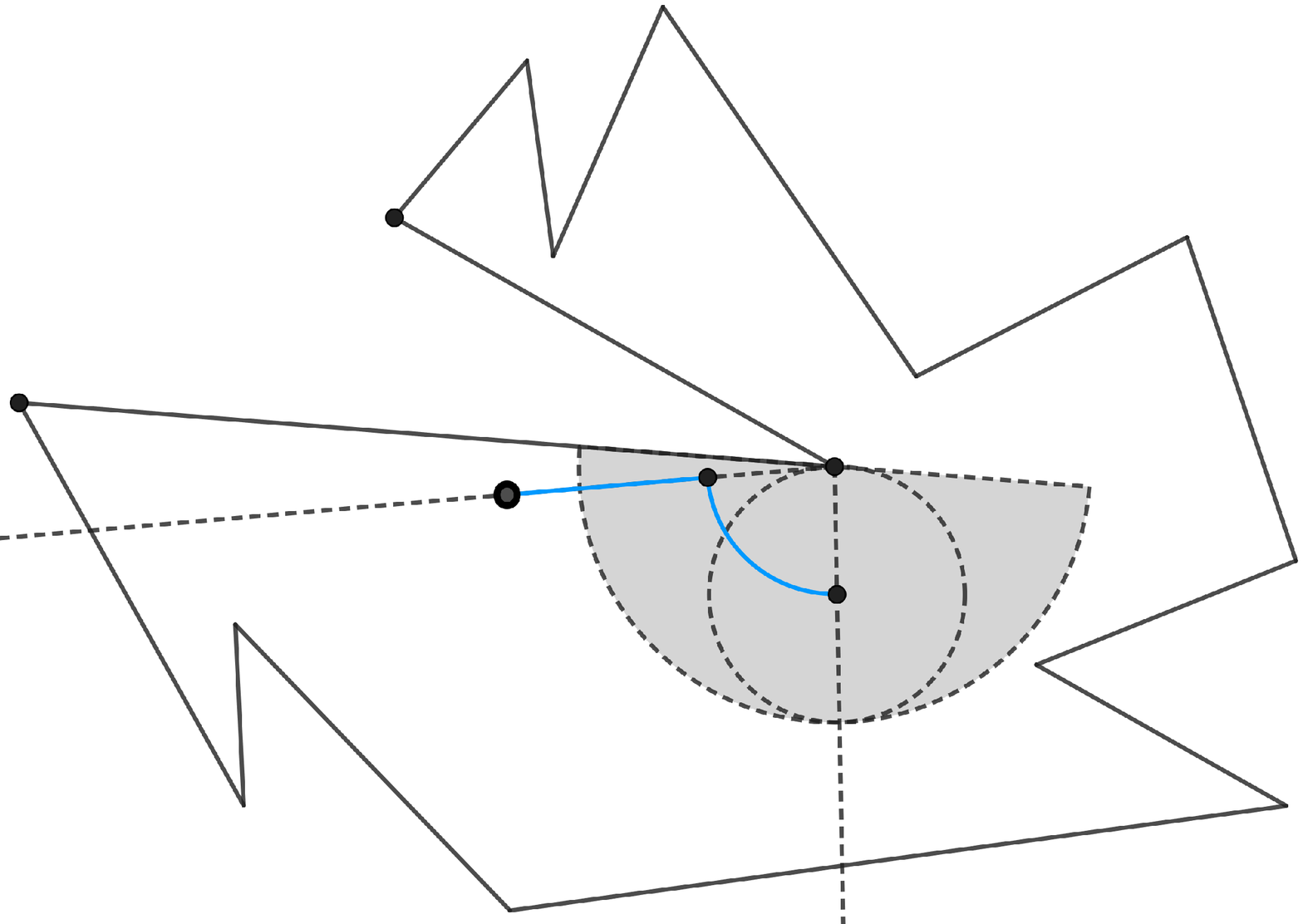
}
\hfill
\subcaptionbox[Short Subcaption]{
     \label{fig: stage2_robot_agree}
}
[
    0.49\textwidth 
]
{
    \fontsize{10pt}{10pt}\selectfont
    \def\svgwidth{0.49\textwidth}
    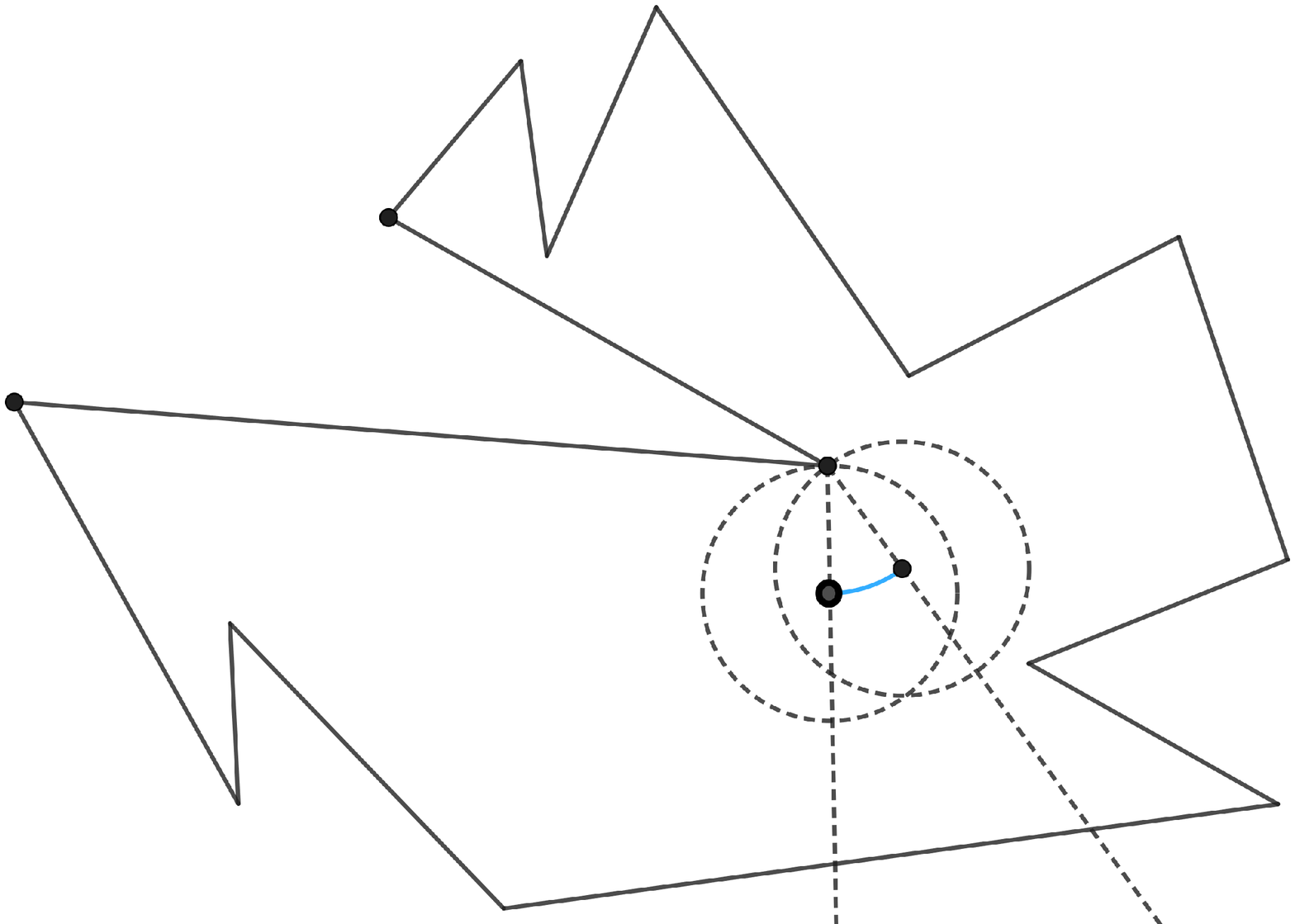
}
\\
\subcaptionbox[Short Subcaption]{
     \label{fig: stage2_target0}
}
[
    0.49\textwidth 
]
{
    \fontsize{10pt}{10pt}\selectfont
    \def\svgwidth{0.49\textwidth}
    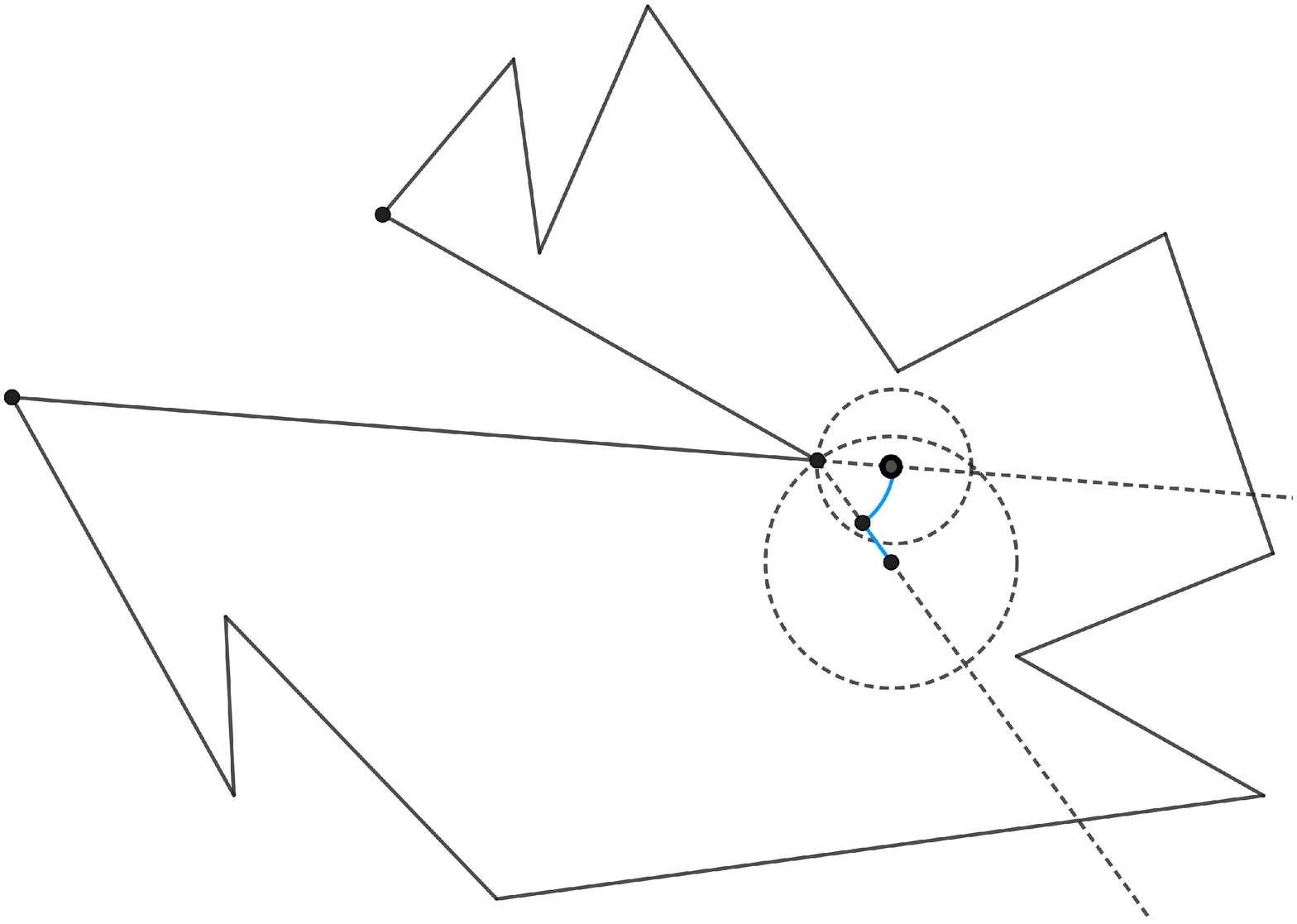
}
\hfill
\subcaptionbox[Short Subcaption]{
     \label{fig: stage2_target0}
}
[
    0.49\textwidth 
]
{
    \fontsize{10pt}{10pt}\selectfont
    \def\svgwidth{0.49\textwidth}
    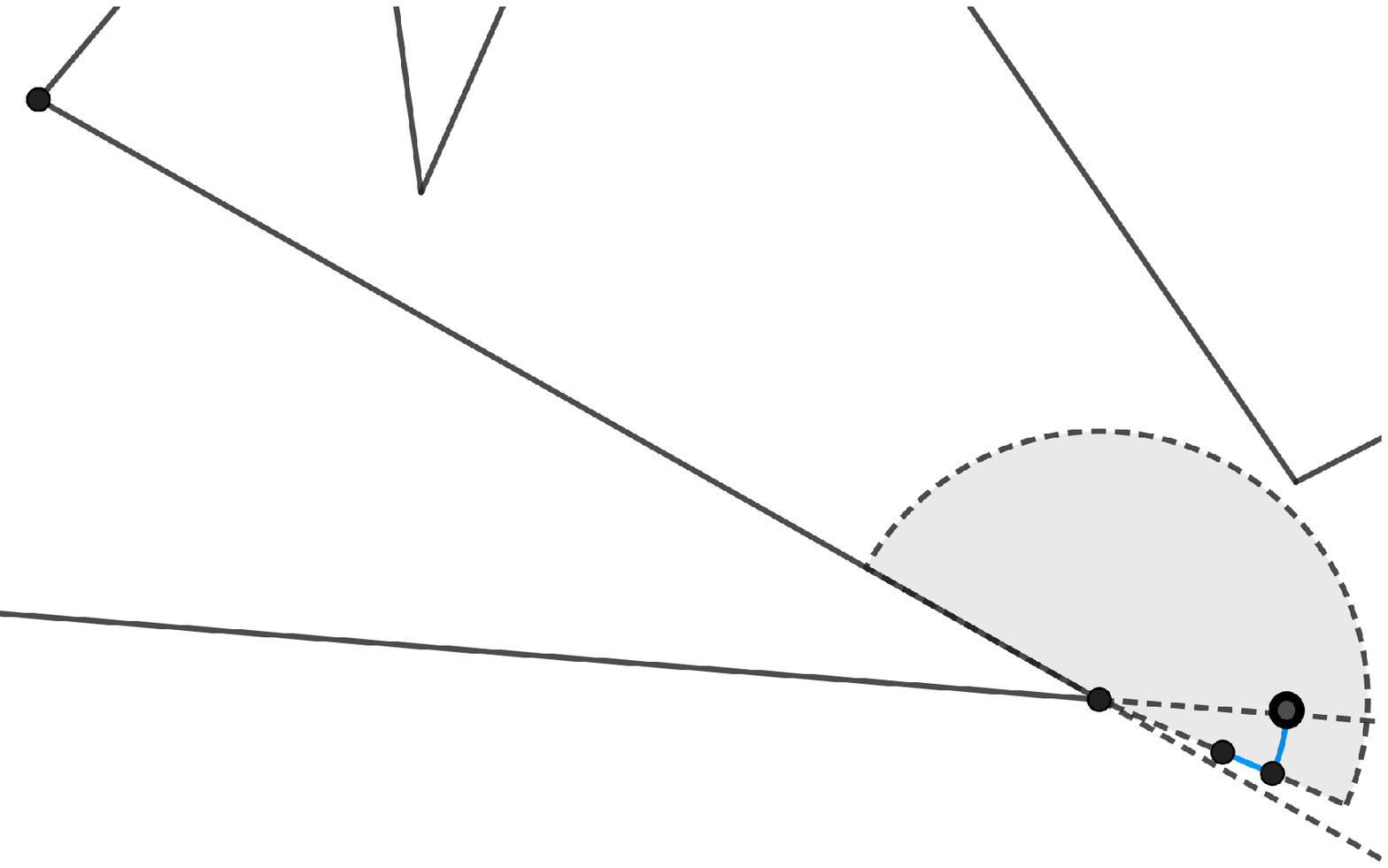
}
\caption[Short Caption]{The robot moving around $p_{i+1}$ to discover the succeeding Vertex and encode a new snapshot. The trail of the robot is shown in blue. }
\label{fig: teq2}
\end{figure}

Now consider the case where $\overline{p_{i+1}p_{i+2}} \setminus \{p_{i+1}\}$ is completely invisible to $r$ (See Fig. \ref{fig: teq2}). This is possible only if $\angle p_ip_{i+1}p_{i+2} > \pi$. Then no matter how close $r$ gets to $p_{i+1}$, $\overline{p_{i+1}p_{i+2}} \setminus \{p_{i+1}\}$ will remain completely invisible to it. In this case, $r$ will move to a point $x''$ that is close enough to $p_{i+1}$, such that the following conditions are satisfied. 

\begin{description}
 \item [D1] $d(x'',p_{i+1}) = d$ should encode all the old data encoded by $\angle x''p_{i+1}p_i$ (both) expressed in the coordinate system  $(p_{i},p_{i+1})$.
   
 \item [D2] Let $S$ be the semicircular disc of radius $2d$, centered at $p_{i+1}$ and having diameter along the line $\overleftrightarrow{p_{i}p_{i+1}}$. $S$ should not intersect with any portion of $\partial(P)$ except $\overline{p_{i}p_{i+1}}$. 
 
 \item [D3] Every point on $\overline{p_{i}p_{i+1}}$ should be visible from every point on $arc(u,\pi,p_{i+1})$, where $u \in \overline{p_{i}p_{i+1}}$ with $d(u,p_{i+1}) = d$.

 \end{description}

   When these conditions are satisfied, the robot will change its state to $s_2$. Clearly, $p_{i+1}$ is its virtual vertex. Let $y$ be a point on the line through $p_{i+1}$ and perpendicular to $\overline{p_{i}p_{i+1}}$, with $d(y,p_{i+1}) = d(x'',p_{i+1}) = d$. The robot will then move to the point $y$ along $arc(x'',y,p_{i+1})$. It implies from condition D2 that as $r$ traverses along this arc (where it can be stopped several times by the adversary), $p_{i+1}$ will remain its virtual vertex. Upon reaching the point $y$, $\overline{p_{i+1}p_{i+2}}\setminus \{p_{i+1}\}$ may still be completely invisible. In that case, $r$ will have to move further along a circular arc and place itself on the extension of the segment $\overline{p_{i}p_{i+1}}$. But if $r$ revolves with the same radius, its virtual vertex may change. Therefore it has to first reduce its distance from $p_{i+1}$. But recall that its distance from $p_{i+1}$ is encoding its memory and hence, the data will be lost if this distance is changed. Therefore, before changing its distance from $p_{i+1}$, it will encode the data `somewhere' else, such that it is preserved while it moves towards $p_{i+1}$. Notice that although moving around $p_{i+1}$ with the same radius can change its virtual vertex, it can still move by a small enough angle without changing its virtual vertex. From its view from $y$, it can compute a point $y''$, such that the following conditions are satisfied.
   
   \begin{description}
    \item [E1] $d(p_{i+1}, y'') = d(p_{i+1}, y) = d$.
    
    \item [E2] $D(y'',d) \cap \partial(P) = \{p_{i+1}\}$. 
    
    \item [E3] $\angle y''p_{i+1}y < \frac{\pi}{2}$ encodes the same data encoded by $d$ both expressed in the coordinate system  $(p_{i},p_{i+1})$. 
   \end{description}

  Now $r$ will first move to $y''$ along a circular arc and then change its state to $s_4$. Then it will reduce its distance from $p_{i+1}$ to $d'$, so that $d'$ satisfies the following conditions. 
  
  \begin{description}
   \item [F1] $d'$ encodes the same data encoded by $\angle y''p_{i+1}y$ both expressed in the coordinate system defined by $(p_{i}p_{i+1})$. 
   
   \item [F2] Let $z$ be the point on the extension of the segment $\overline{p_{i}p_{i+1}}$ with $d(z,p_{i+1})=d'$. Then $D(z,d') \cap \partial(P) = \{p_{i+1}\}$. 
  \end{description}

  When these conditions are satisfied, it will change its state to $s_2$. Now $r$ will move to $z$ by moving around $p_{i+1}$ in counterclockwise direction maintaining the distance $d'$ from it. Upon reaching $z$, it can see at least some portion of $\overline{p_{i+1}p_{i+2}} \setminus \{p_{i+1}\}$. Suppose that it still can not see $p_{i+2}$. But since it can see some portion of $\overline{p_{i+1}p_{i+2}} \setminus \{p_{i+1}\}$, it can compute the point $z'$ on the extension of the segment $\overline{p_{i+2}p_{i+1}}$ with $d(z',p_{i+1})=d'$. Now $r$ will move around $p_{i+1}$ in clockwise direction towards $z'$, but not touching it (say by choosing the middle of $arc(z',z,p_{i+1})$ as its destination and so on). Eventually, it will be able to see $p_{i+2}$. In fact, it can see both $\overline{p_{i+1}p_{i+2}}$ and $\overline{p_{i}p_{i+1}}$ entirely. Now $r$ has to encode a new snapshot (merged with the old ones) in its distance from $p_{i+1}$. Before that it will encode its memory in the angle that it makes with the extension of $\overline{p_{i+2}p_{i+1}}$ at $p_{i+1}$ by revolving further towards $z'$, and then will change its state to $s_5$. Then it will move towards $p_{i+1}$ so that conditions C1, C2 are satisfied. When they are achieved, $r$ will change its state to $s_1$. 
  
\subsubsection{Taking a Second Tour of a Connected Component of the Boundary}\label{teq3}

From the techniques discussed in the Sections \ref{teq1} and \ref{teq2}, it is clear how a robot can `visit' all the vertices of a previously unexplored connected component $C$ of $\partial(P)$. Also, whenever $r$ encodes a new snapshot, it marks the position of its current virtual vertex with a `visited once' flag. Upon completing its first tour of $C$, it will start a second tour of $C$ in the same direction. In the second tour, the points from where the snapshots are taken, should constitute an `approximation' of $C$, say $\overline{C}$, such that the closed polygonal curve $\overline{C}$ 1) does not self-intersect, 2) does not intersect $\partial(P)$, and 3) does not intersect any other previous approximations.

\begin{figure}[h!]
\subcaptionbox[Short Subcaption]{
       \label{fig: dA}
}
[
    0.45\textwidth 
]
{
    \fontsize{10pt}{10pt}\selectfont
    \def\svgwidth{0.45\textwidth}
    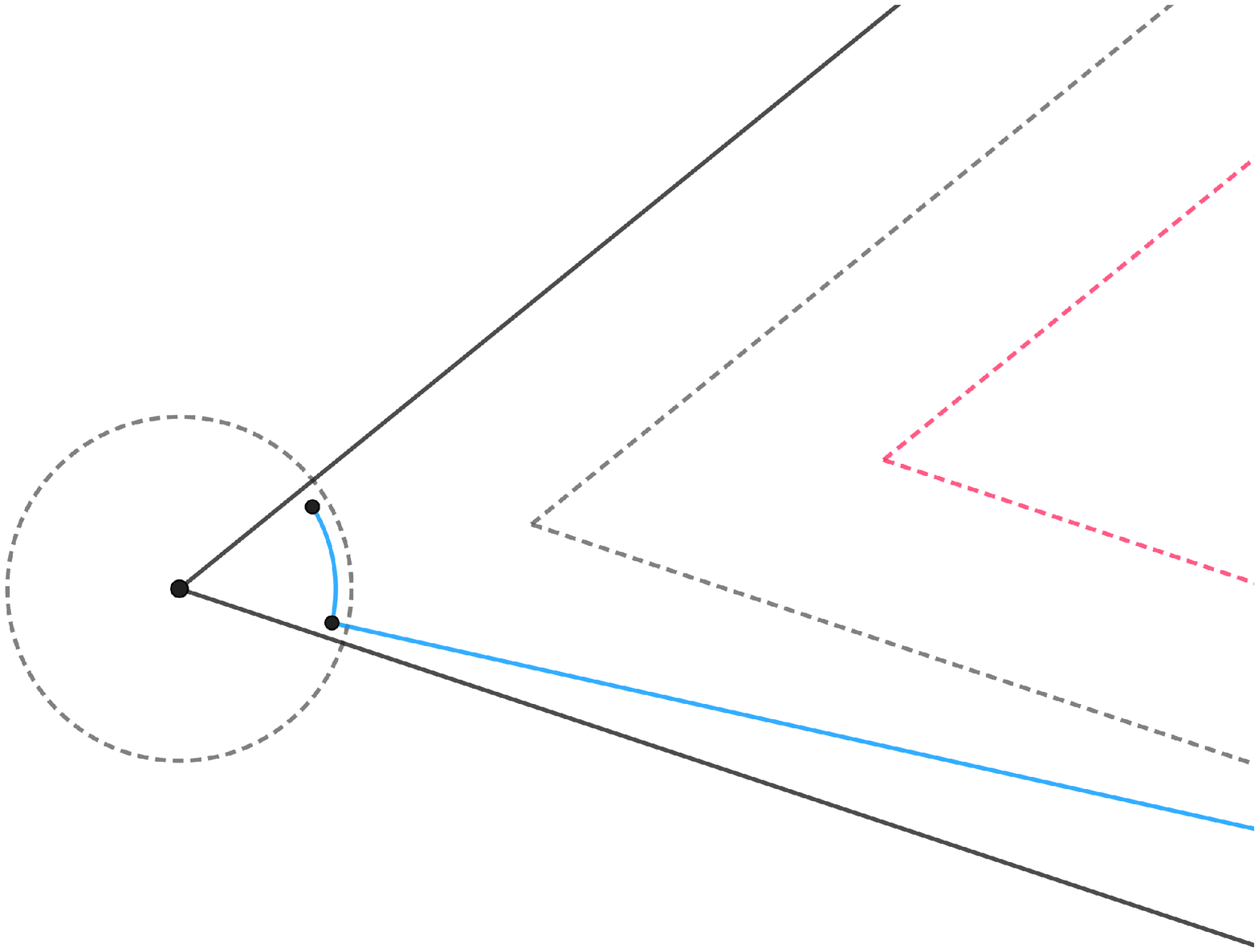
}
\hfill
\subcaptionbox[Short Subcaption]{
     \label{fig: gfB}
}
[
    0.45\textwidth 
]
{
    \fontsize{10pt}{10pt}\selectfont
    \def\svgwidth{0.45\textwidth}
    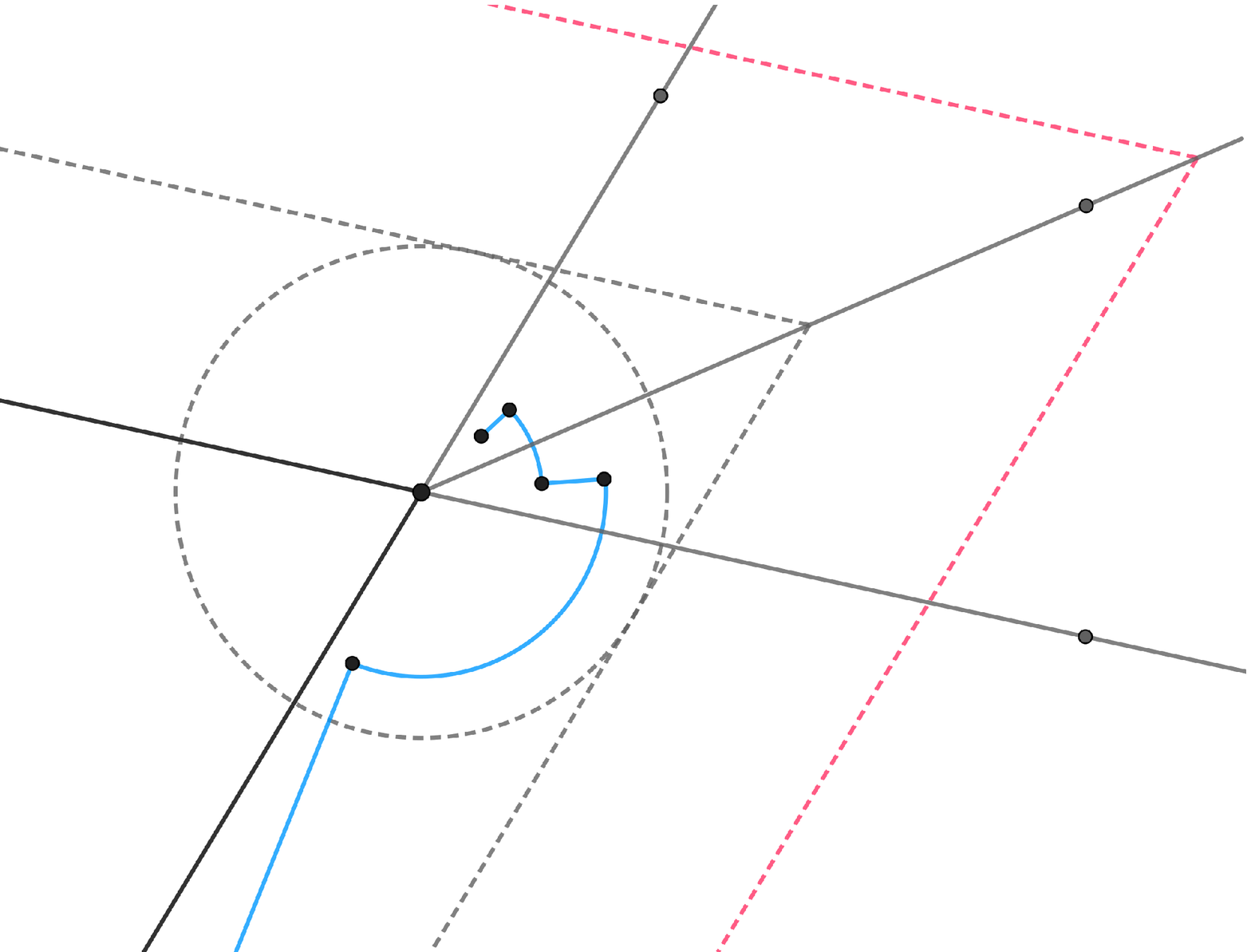
}
\caption[Short Caption]{The robot taking a second tour. The trail of the robot is shown in blue. The approximations $\tilde{C}(d)$ and $\tilde{C}({\frac{d}{2}})$ are shown in pink and grey dotted lines respectively. }
\label{fig: tour2}
\end{figure}

Suppose that $C$ is composed of $m$ vertices $p_1, \ldots, p_m$. Assume that $r$ has started exploring $C$ from (close to) $p_1$. As described earlier, it will sequentially visit all the vertices and eventually arrive at a point close to $p_m$, from where $p_1$ is visible. It can clearly identify $p_1$ to be a previously visited vertex and will decide to start the second tour. Now clearly $r$ has a full picture of $C$. So it can compute a distance $d$ implicitly and include it in its memory, so that $d$ has the following property. Let $\tilde{C}(d) = \{{p_1'},\ldots,{p_m'}\}$ denote the approximation of ${C} = \{{p_1},\ldots,{p_m}\}$ such that each $\overline{{p_i'}{p_{i+1}'}}$ is parallel to $\overline{p_ip_{i+1}}$ ($p_{m+1}$ is to be understood as $p_1$) and the separation between them is $d$ (See Fig. \ref{fig: tour2}). Then $d$ should be small enough, so that the approximation $\tilde{C}(d)$ satisfies all the three requirements.  The points from where the robot will take snapshots during its second tour, will constitute an approximation $\overline{C} = \{p^1_1, p^2_1,\ldots p^1_m, p^2_m\}$ consisting of $2m$ points, with $\overline{C}$ lying in the region between $C$ and $\tilde{C}(d)$. We shall now discuss the procedure in detail. The robot will approach $p_1$ (with state $s_3$) in the same manner as described previously, but with an extra requirement that the path it follows should be lying in the region between $C$ and $\tilde{C}({\frac{d}{2}})$. Note that although $d$ is computed in the implicit form, $r$ can get an approximation of $d$ in explicit form that is smaller than the actual value. Now there are two cases to consider. 

First consider the case where $\angle p_mp_1p_2$ is not reflex. Similar to the first tour, $r$ goes to a point $x$ so that the conditions C1 and C2 are satisfied (with $p_{i} = p_m, p_{i+1} = p_1, p_{i+2} = p_2$). We can refer to this in short by simply saying that `$r$ takes a snapshot at $x$'. The extra requirement in this case would be that $d(x,p_1) < \frac{d}{2}$. After this, $r$ will change its state to $s_1$. Note that from the encoded data, the robot understands that it is currently taking a second tour of $C$. Now $r$ will move around $p_1$ to reach a point $x'$ so that the condition  B2 (with $p_i = p_1, p_{i+1} = p_2$) is satisfied, plus $\angle x'p_1p_2$ should encode the view from $x'$ merged with the older snapshots (encoded by $d(x',p_1)$) expressed in the coordinate system $(p_1,p_2)$. Again using similar phrasing, we shall refer to this by saying `$r$ takes a snapshot at $x'$'. Let us denote the points $x$ and $x'$ by $p^1_1$ and $p^2_1$. Note that our constructions ensure that the line segment $\overline{p^1_1p^2_1}$ is inside the region $C$ and $\tilde{C}({d})$. 

%

Now consider the case where $\angle p_mp_1p_2$ is reflex. The robot will go to a point $x$ that is close enough to $p_1$, such that the following conditions are satisfied.

\begin{description}

%
 \item [G1] $d(x,p_1)$ encodes the old snapshots (encoded in $\angle xp_1p_m$) expressed in the coordinate system defined by $(p_m,p_1)$.
 
 \item [G2] $d(x,p_1) < \frac{d}{2}$.
 
\end{description}
  
After reaching such a point $x$, $r$ will change its state to $s_2$. Let $\overrightarrow{p_1A}$ and $\overrightarrow{p_1B}$ be the extensions of the segments $\overline{p_2p_1}$ and $\overline{p_mp_1}$ respectively. Let $\overrightarrow{p_1O}$ be the angular bisector of the angle $\angle Ap_1B$. Now $r$ can move around $p_1$ to place itself at a point $p^1_1$ between the lines $\overrightarrow{p_1A}$ and $\overrightarrow{p_1O}$ such that the angle $\angle p^1_1p_1O$ encodes the view from $p^1_1$ merged with the older snapshots all expressed in the coordinate system defined by $(p_m,p_1)$. In other words, $r$ takes a snapshot at $p^1_1$. Then $r$ will change its state to $s_6$, move towards $p_1$ to encode the data in its distance from $p_1$, again change its state to $s_2$ and move around $p_1$ to take a snapshot at a point $p^2_1$ between the lines $\overrightarrow{p_1O}$ and $\overrightarrow{p_1B}$ encoding the snapshot (merged with the old ones) in the angle $\angle p^2_1p_1O$ expressed in the coordinate system defined by $(p_m,p_1)$. Then $r$ will again change its state to $s_6$ and move towards $p_1$ to encode the data in its distance from $p_1$, this time expressed in the coordinate system $(p_1,p_2)$. After this, it will change its state to $s_1$. Continuing in this manner, the robot will revisit all the vertices of the component, and take snapshots at $p^1_i$ and $p^2_i$, near each vertex $p_i$. The polygonal chain $\overline{C}$ clearly satisfies all three desired properties.

%

\subsubsection{Moving from One Connected Component to Another}

A robot will move from a virtual vertex $p_i$ to a vertex $p_j$ belonging to a different connected component of $\partial (P)$ only if $\overline{p_ip_j} \subset PVor_P(p_i) \cup PVor_P(p_j)$. The robot $r$ with state $s_3$ will approach $p_i$ and encode its memory in its distance from $p_i$ expressed in the coordinate system defined by $(p_{i-1},p_{i})$. The robot will then change its state to $s_2$.  Note that $p_j$ may not even be visible from its current position if $\angle p_{i-1}p_ip_{i+1}$ is reflex. If $\angle p_{i-1}p_ip_{i+1}$ is reflex and $p_j$ lies in the open half-plane delimited by $\overleftrightarrow{p_{i-1}p_i}$ containing $p_{i+1}$, it will have to encode its memory in the coordinate system $(p_{i},p_{i+1})$ by previously discussed techniques. It will then change its state to $s_1$. From its memory, it knows that the plan is to move to $p_j$. It will then move around $p_i$ to move to a point $x$ so that the following conditions are satisfied.

%
%

\begin{description}

 \item [H1] The ray $\overrightarrow{p_ix}$ intersects the interior of the Voronoi edge $Vor_{S(x)}(p_i) \cap Vor_{S(x)}(p_j)$, where $S(x)$ denotes the polygon vertices visible from $x$. Suppose that the ray intersects the  Voronoi edge $Vor_{S(x)}(p_i) \cap Vor_{S(x)}(p_j)$ at point $A$.   
 
 \item [H2] The angle $\alpha = \angle xp_ip_j$ encodes its memory. All coordinates of the snapshots are expressed in the coordinate system defined by $(p_i,p_j)$. The encoding will also contain a rational approximation of $\frac{1}{2}(p_i - p_j)$ expressed in the local coordinate system of $r$.
\end{description}


\begin{floatingfigure}[r]{7.0cm}
    \fontsize{10pt}{10pt}\selectfont
    \def\svgwidth{0.35\textwidth}
\begingroup%
  \makeatletter%
  \providecommand\color[2][]{%
    \errmessage{(Inkscape) Color is used for the text in Inkscape, but the package 'color.sty' is not loaded}%
    \renewcommand\color[2][]{}%
  }%
  \providecommand\transparent[1]{%
    \errmessage{(Inkscape) Transparency is used (non-zero) for the text in Inkscape, but the package 'transparent.sty' is not loaded}%
    \renewcommand\transparent[1]{}%
  }%
  \providecommand\rotatebox[2]{#2}%
  \ifx\svgwidth\undefined%
    \setlength{\unitlength}{396bp}%
    \ifx\svgscale\undefined%
      \relax%
    \else%
      \setlength{\unitlength}{\unitlength * \real{\svgscale}}%
    \fi%
  \else%
    \setlength{\unitlength}{\svgwidth}%
  \fi%
  \global\let\svgwidth\undefined%
  \global\let\svgscale\undefined%
  \makeatother%
  \begin{picture}(1,0.92323232)%
    \put(0,0){\includegraphics[width=\unitlength]{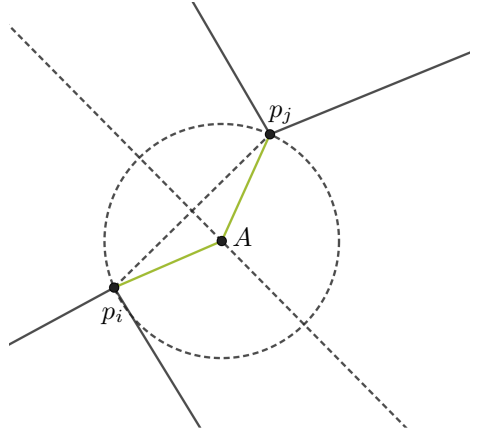}}%
    \put(0.20135251,0.23844376){\color[rgb]{0,0,0}\makebox(0,0)[lb]{\smash{$p_i$}}}%
    \put(0.56404715,0.67700621){\color[rgb]{0,0,0}\makebox(0,0)[lb]{\smash{$p_j$}}}%
    \put(0.48443323,0.39609967){\color[rgb]{0,0,0}\makebox(0,0)[lb]{\smash{$A$}}}%
  \end{picture}%
\endgroup%

\caption{$p_i$ and $p_j$ are vertices belonging to different connected components of $\partial(P)$. The robot will move along the path $\mathcal{P}(p_i,p_j,\alpha)$ drawn in green.}
\label{phase4}
\end{floatingfigure}

The robot will then change its state to $s_7$, move along $\mathcal{P}(p_i,p_j,\alpha)$ towards $p_j$, i.e., it will first move to $A$ and then to a point properly close to $p_j$. Consider the situation when $r$ stops at a point $z$ on the path $\mathcal{P}({p_i,p_j,\alpha})$. When $r$ was at $x$, it verified H1 by checking from its snapshot that the disc $D(A, d)$ contains no polygon vertex other than $p_i, p_j$, where $d = d(A,p_i) = d(A,p_j)$. It implies from this that $\mathcal{P}({p_i,p_j,\alpha}) \subset PVor_P(p_i) \cup PVor_P(p_j)$.  Hence, at $z$, its closest visible vertex, and hence its virtual vertex, is either $p_i$ or $p_j$. It computes intersections between the ray from its virtual vertex, passing through it, and the perpendicular bisectors of the lines joining its virtual vertex and other visible vertices; and then checks if the intersection point is on the corresponding Voronoi edge. It will find that the ray intersects the Voronoi edge defined by $p_i$ and $p_j$ first. However, $r$ does not immediately know whether it is moving from $p_i$ to $p_j$, or from $p_j$ to $p_i$. However, it knows that its memory is encoded in the angle it makes with $\overline{p_ip_j}$ at its virtual vertex $\in \{p_i,p_j\}$. But $r$ does not to know if it is encoded with the coordinate system $(p_i,p_j)$ or $(p_j,p_i)$. However, recall that the memory contains a rational approximation of $\frac{1}{2}(p_i - p_j)$, call it $w$, expressed in its local coordinate system. Now $r$ computes $w + \frac{1}{2}(p_i + p_j)$, which gives an approximation of $p_i$.  from which $r$ determines that it is moving away from $p_i$, and also the fact that its encoded memory is expressed in $(p_i,p_j)$. So, eventually it will move to a point properly close to $p_j$ so that its distance from $p_j$ encodes its memory expressed in either $(p_j,p_{j+1})$ or $(p_{j-1},p_j)$, and then change its state to $s_1$ or $s_2$ accordingly.

%

 \subsection{The Main Result}\label{result}

We shall assume that at the beginning, either the robot is at a polygon vertex with any arbitrary internal state, or the robot is anywhere inside the polygon, but its internal state is set to a special value $s_0$. In the later case, the robot can move to the closest polygon vertex. Therefore, we can assume that the starting position of the robot $r$ is at a polygon vertex $p_0$, and its internal state is arbitrary.

When $r$ is first activated, it will take a snapshot and compute a path $\mathcal{P}(p_0,p_{1},\alpha)$ such that $\mathcal{P}(p_0,p_{1},\alpha)$ is inside $LVor_{E(P)}(\overline{p_0p_{1}})$, and $\alpha$ encodes the snapshot in the coordinate system  $(p_{0},p_{1})$. Then it will change its state to $s_3$ and start moving along the path. Then it will explore the connected component of $\partial(P)$ containing $p_0$ twice as discussed in Section \ref{teq1}, \ref{teq2} and \ref{teq3}. 

Then $r$ will have to go to an unvisited vertex of a different connected component. Note that a robot moves from a vertex $p_i$ to a vertex $p_j$ belonging to a different connected component only if $\overline{p_ip_j} \subset PVor_P(p_i) \cup PVor_P(p_j)$. We show that there  will be always possible to reach an unvisited vertex (of a different connected component) respecting this rule. Let $V$ and $U$ be respectively the set of visited and unvisited vertices of $P$ till now. Let $(p_i, p_j) \in V \times U$ be the closest pair of mutually visible vertices from the two sets. Let $d(p_i, p_j) = d$. Then the interior of $D(p_i, d)$ contains no vertex of $U$, other than $p_j$, visible from $p_i$. Similarly, the interior of $D(p_j, d)$ contains no vertex of $V$, other than $p_i$, visible from $p_j$. It is not difficult to see that if $D(p_i, d) \cap D(p_j, d)$ contains some vertex $p$ other than $p_i$ and $p_j$, then there is a boundary edge that intersects the boundary of $D(p_i, d) \cap D(p_j, d)$ at two points and separates $p$ from $p_i$ and $p_j$. Take any point $z \in \overline{p_ip_j}$, and let $d' = min\{d(z, p_i), d(z, p_j)\}$. Since $D(z, d') \subset D(p_i, d) \cap D(p_j, d)$, there is no polygon vertex in $D(z, d')$ that is visibile from $z$ other than $p_i$ or $p_j$ or both. So, $z \in PVor_P(p_i) \cup PVor_P(p_j)$. Therefore, we have $\overline{p_ip_j} \subset PVor_P(p_i) \cup PVor_P(p_j)$. Since $p_i$ is a visited vertex, there must be a legal path to $p_i$ from the current position of the robot. Therefore, there exists a legal path to $p_j$ from the current position of the robot.

As mentioned earlier, whenever the robot goes to a previously unexplored connected component of $\partial(P)$, it will visit all its vertices twice. It will repeat this process until there are no unvisited vertices left in its memory. We show that eventually all vertices of $P$ are indeed visited. The arguements are similar to \cite{Luna17}. The procedure ends when there are no vertices of $P$ left in the encoded memory of $r$ that are marked `unvisited'. Hence, any vertex of $P$ is either undiscovered or `visited'. Let $U \subset V(P)$ be the set of vertices that are undiscovered. We have to show that $U = \emptyset$. When the procedure ends, $r$ has touched all the vertices of some approximation $\overline{C}$ of $P$, which are mutually disjoint simple closed polygonal chains $\overline{C_j}$ which approximate the components $C_j$ of $\partial(P)$. Let us construct a new polygon $P'$ by removing every $C_j$ from the boundary of $P$ and replacing it with $\overline{C_j}$. $P'$ is a polygon because of the way the $\overline{C_j}$ 's have been constructed. Therefore, the visibility graph of $V(P')$ is connected.  Notice that the vertices in $U$ are also vertices of $P'$, i.e., $U \subset P'$. Suppose that $U \neq \emptyset$. Now consider the visibility graph $G$ of $P'$. Since $G$ is connected, there is a non-empty cut-set of $U$ and $V(P') \setminus U$ in $G$. Take any edge $e = (u,v)$ from the cut-set, where $u \in U$ and $v \in V(P') \setminus U$. Then $u$ is visibile from $v$. Recall that $r$ has touched every vertex of $V(P') \setminus U$ and taken snapshots from there. Hence, $r$ has taken a snapshot exactly from the point $v$, and therefore, must have seen $u$. This contradicts the fact that $u \in U$. 

Therefore, we can conclude the main result the paper as the following.
 
\begin{theorem}\label{thm main}
 In $\textsc{FState}$, a robot inside a polygon $P$ with non-rigid movements can correctly construct and encode a map of the polygon in finite time.
\end{theorem}

\section{Conclusion}\label{conclu}

In this work, we have shown how a finite state robot with non-rigid movements can construct the map of a polygon by a positional encoding strategy.  The techniques developed here, give a general movement strategy for finite state robots with non-rigid movements, to move about in the polygon, without losing its encoded memory. The map construction algorithm can be used as a subroutine to solve distributed algorithms for mobile robot systems under this model, where the knowledge of the polygon may be required.  For instance, consider the \textsc{Gathering} problem, where a set of autonomous, anonymous, asynchronous finite state mobile robots with no agreement in coordinate system and no communication capabilities, have to meet at some point in the polygon. Assume that the polygon is asymmetric. Then each robot will first construct and encode the map of the polygon. Since the polygon is asymmetric, the robots can deterministically pick a polygon vertex as their meeting point. Then using our techniques, the robots can move to that vertex. However, when the polygon is not asymmetric, \textsc{Gathering} appears to challenging even for robots with unlimited memory. For symmetric polygons, we can consider the relaxed version of  \textsc{Gathering}, called \textsc{Meeting}, where any two of the robots have to become mutually aware by seeing each other at their LOOK phases. Using our techniques, a patrolling strategy similar to \cite{Luna17} can be adapted to our setting to solve \textsc{Meeting}.

It would be very interesting to investigate whether map construction or \textsc{Meeting} can be solved by fully oblivious robots with non-rigid movements. Another direction would be to study the problems for oblivious robots with limited visibility. Also, our movement model allows the robots to make circular moves, as opposed to \cite{Luna17}, where the robots can move only along a straight line. The ability to make circular moves is crucial to our algorithm. Although the standard model for mobile robots in the plane assumes rectilinear robot movement, it is not completely uncommon to allow the robots to move along circular trajectories (e.g., \cite{Cicerone17, Fujinaga15}). It would be interesting to see if the same result can be achieved without the ability to make circular moves.

 \paragraph{Acknowledgements.} The first three authors are supported by NBHM, DAE, Govt. of India, CSIR, Govt. of India and UGC, Govt. of India, respectively. We would like to thank the anonymous reviewers for their valuable comments which helped us improve the quality and presentation of the paper.

\bibliographystyle{plain}
\bibliography{encoding_polygon}

\end{document}